\documentclass[12pt]{article}

\usepackage{amsmath}
\usepackage{amssymb}
\usepackage{latexsym}
\usepackage{graphics}
\usepackage{psfrag,fancyhdr,epsfig}

\usepackage{color}

\definecolor{orange}{rgb}{0.9,0.2,0}
\definecolor{brown}{rgb}{0.7,0.3,0.2}
\definecolor{fuxia}{rgb}{1,0,1}
\definecolor{skyblue}{rgb}{0,0.1,0.9}
\definecolor{violetred}{rgb}{0.8,0.13,0.56}
\definecolor{deeppink}{rgb}{1.00,0.08,0.5}
\definecolor{pink}{rgb}{1.00,0.75,0.80}
\definecolor{orchid}{rgb}{0.85,0.44,0.84}
\definecolor{lightpink}{rgb}{1.00,0.71,0.76}
\definecolor{bluish}{rgb}{0,0.6,0.8}

\begin{document}
\begin{titlepage}

\begin{centering}

\hfill hep-ph/yymmnnn\\

\vspace{1 in}
{\bf {On Discrete R-Symmetries in MSSM and its Extensions} }\\
\vspace{1 cm}
{M. Paraskevas and K. Tamvakis}\\
\vskip 0.5cm
{$ $\it{Physics Department, University of Ioannina\\
GR451 10 Ioannina, Greece}}

\vskip 0.5cm

\vspace{1.5cm}
{\bf Abstract}\\
\end{centering}
\vspace{.1in}

 We study possible anomaly-free discrete $R$-symmetries that avoid the $\mu$-problem and the dangerous $D\leq 5$ operators considering charge assignments that do not commute with the traditional grand unifying simple groups, such as $SU(5)$ or $SO(10)$, but commute instead with the so-called flipped-$SU(5)$, with or without the operation of the GS mechanism. We find ${\cal{Z}}^{(R)}_N$ symmetries with $N=3,\,6$ in the anomaly-free case or $N=3,\,4,\,6,\,8,\,12,\,24$ in the case of anomaly cancellation through the Green-Schwartz mechanism. Non-unified cases (${\cal{Z}}_5$) have also been discussed. We also confront the construction of a $4D$ grand unified flipped-$SU(5)$ model endowed with ${\cal{Z}}^{(R)}_N$ and find phenomenologically acceptable solutions with  $N\,=\,2k+7$ and $N\,=\,2k+10$.
 \vfill

\vspace{2cm}
\begin{flushleft}

May 2012
\end{flushleft}
\hrule width 6.7cm \vskip.1mm{\small \small}
 \end{titlepage}

\section{ Introduction and General Framework}
Supersymmetric unified models, such as MSSM or extensions of it, have been proposed mainly as a framework that eliminates, at least, the technical aspects of the hierarchy problem. Apart from that and various other attractive ingredients, such as the unification of gauge couplings and the, by default, existence of dark matter candidates, they have a number of problems of their own. These are the so-called $\mu$-problem and the need to eliminate dangerous $D\leq 5$ Baryon/Lepton violating operators. {\textit{Discrete symmetries}} have been introduced as a means to control these potential problems\cite{FF, DRW, IR, BGW, DLT}. In addition to that, discrete symmetries, and in particular {\textit{discrete $R$-symmetries}} might play a role with respect to the scale of supersymmetry breaking, parametrized by the gravitino mass, since this can be controlled by existing discrete $R$-symmetries of the superpotential. Therefore, if supersymmetry is going to be interesting for low energy physics, it should be important to investigate the role of such discrete $R$-symmetries. Such a discrete symmetry has been shown to avoid the $\mu$-problem and protect the electroweak scale in singlet extensions of MSSM\cite{PT}. Nevertheless, there exist convincing arguments that a discrete symmetry should be embeddable in an anomaly-free gauge symmetry, otherwise quantum gravitational effects would violate it\cite{KW, BD}. Thus, an anomaly-free discrete $R$-symmetry should be employed. Cancellation of anomalies may be direct\cite{EIKY} or proceed through the {\textit{Green-Schwarz mechanism (GS)}}. Such a discrete $R$-symmetry, commuting with $SO(10)$, was shown to avoid the $\mu$-problem in MSSM and proton decay through $D=5$ operators\cite{BGW-1}\cite{LRRRSHV}. In the present article we extend the investigation of possible anomaly-free discrete $R$-symmetries that avoid the $\mu$-problem and the dangerous $D\leq 5$ operators by considering charge assignments that do not commute with the traditional grand unifying simple groups, such as $SU(5)$ or $SO(10)$, but commute instead with the so-called flipped-$SU(5)$, with or without the operation of the GS mechanism. We stay within the framework of $MSSM$, discussing briefly possible singlet or multidoublet extensions. We find that, in the anomaly free case, the symmetries ${\cal{Z}}_3^{(R)},\,{\cal{Z}}_6^{(R)}$ with flipped assignments are possible and present sets of phenomenologically acceptable charges. In the case of GS-anomaly cancellation, we recover ${\cal{Z}}_N^{(R)}$ symmetries with flipped assignments for $N\,=\,3,\,4,\,6,\,8,\,12,\,24$ and present sets of phenomenologically acceptable charges. In addition, we discuss a singlet extension of $MSSM$ with ${\cal{Z}}_5^{(R)}$ symmetry as an example of non-unified charge assignments. Finally, we confront the issue of constructing a $4D$ grand unified theory endowed with such symmetry and arrive at an extended flipped $SU(5)\times U(1)_X$ model endowed with an anomaly free (through GS) symmetry ${\cal{Z}}_N^{(R)}$ for $N\,=7$ and $N\,\geq 9$. Sets of phenomenologically allowed charge assignments are listed.

Next, we proceed to review briefly the general framework. Consider an $R$-symmetry ${\cal{Z}}_N$, under which chiral superfields transform as $\Phi\rightarrow\,e^{i\frac{2\pi q}{N}}\,\Phi$, where the integer $q$ is the {\textit{charge}} of $\Phi$ under it. We are free to choose the superpotential charge to have a specific value and we take $q_{{\cal{W}}}\,=\,2$. The {\textit{anomaly coefficients}} corresponding to  $(gauge)^2{\cal{Z}}_N$ and $(grav)^2{\cal{Z}}_N$ anomalies, denoted respectively by ${\cal{A}}$ and ${\cal{A}}_0$,  are
$${\cal{A}}\,=\,\sum_f\ell_f(q_f-1)\,+\,\ell(adj),\,\,\,\,\,{\cal{A}}_0\,=\,-21\,+\,\sum  \ell(adj)\,+\,\#(U(1))\,+\,\sum_a(q_a-1)\,,$$
where, in ${\cal{A}}$, $\ell_f$ is the Dynkin index of the fermion representation $f$ with ${\cal{Z}}_N$ charge $q_f$, $\ell(adj)$ is the contribution of the gauginos and,  in ${\cal{A}}_0$, $-21$ is the contribution of the gravitino, $\sum  \ell(adj)\,+\,\#(U(1))\,$ is the contribution of the gauginos and $q_a$ sums over all remaining fermion charges. The ${\cal{Z}}_N^3$ anomaly coefficient is not considered, since the corresponding condition can be interpreted as an {\textit{embedding condition}}. In the case of the MSSM, we have
\begin{equation}\begin{array}{l}
{\cal{A}}_3\,=\,3\,+\,\frac{N_f}{2}\left(\,2q_Q+q_{d^c}+q_{u^c}\,-4\right)\\
\,\\
{\cal{A}}_2\,=\,2\,+\,\frac{N_f}{2}\left(\,3q_Q+q_{\ell}\,-4\right)\,+\,\frac{1}{2}(q_{h_{u}}+q_{h_{d}}\,-2)\\
\,\\
{\cal{A}}_1\,=\,\frac{N_f}{10}\left(\,q_Q+2q_{d^c}+8q_{u^c}+3q_{\ell}+6q_{e^c}\,-20\right)\,+\,\frac{3}{10}(q_{h_{u}}+q_{h_{d}}-2)\\
\,\\
{\cal{A}}_0\,=\,-9\,+\,(-1)\,+\,N_f\left(\,6q_Q+3q_{u^c}+{3}q_{d^c}+2q_{\ell}+q_{e^c}\,+\,{q_{N^c}}\,-16\,\right)\\
\\
\,+\,2(q_{h_{u}}+q_{h_{d}}\,-2)\,+\,\sum_s(q_s-1)
\end{array}{\label{ANCOEFGEN}}\end{equation}
In the gravitational coefficient we have included the contribution of the {\textit{right-handed neutrino}} $N_f(q_{N^c}-1)$ and the contribution of the {\textit{dilatino/axino}} $(-1)$. We have also left room for the contribution of an unspecified number of singlets with charges $q_s$.

For  ${\cal{Z}}_N$-charges commuting with $SU(5)$, the coefficients are
\begin{equation}\begin{array}{l}
{\cal{A}}_3\,=\,3\,+\,\frac{N_f}{2}\left(\,3q_{10}\,+\,q_{\overline{5}}\,-4\right)\\
\,\\
{\cal{A}}_2\,=\,2\,+\,\frac{N_f}{2}\left(\,3q_{10}\,+\,q_{\overline{5}}\,-4\right)\,+\,\frac{1}{2}(q_{h_{u}}+q_{h_{d}}\,-2)\\
\,\\
{\cal{A}}_1\,=\,\frac{N_f}{2}\left(\,3q_{10}\,+\,q_{\overline{5}}\,-4\right)\,+\,\frac{3}{10}(q_{h_{u}}+q_{h_{d}}-2)\\
\,\\
{\cal{A}}_0\,=\,-9\,+\,(-1)\,+\,N_f\left(\,10q_{10}\,+\,5q_{\overline{5}}\,+\,q_{1}\,-16\,\right)\\
\\
\,+\,2(q_{h_{u}}+q_{h_{d}}\,-2)\,+\,\sum_s(q_s-1)
\end{array}{\label{ANCOEF}}\end{equation}

\section{Non-anomalous Discrete $\bf{R}$-Symmetries} One possibility to obtain an anomaly-free theory in MSSM is to impose
\begin{equation}{\cal{A}}_3\,=\,{\cal{A}}_2\,=\,{\cal{A}}_1\,=0\,\,\,mod(\eta)\,,\,\end{equation}
where the condition on ${\cal{A}}_1$ is required to be met by at least on shifted set of charges. The parameter $\eta$ is $\eta\,=\,N/2$ for $N=2k$ and $\eta\,=\,N$ for $N=2k+1$. These are considered independently of the gravitational coefficient which, in general, depends to an unknown set of gauge singlets. The conditions, in the case of $SU(5)$-invariant charges ({\ref{ANCOEF}}), are equivalent to
\begin{equation}\begin{array}{l}
{\cal{A}}_3\,=\,0\,\,mod(\eta)\\
\,\\
{\cal{A}}_2\,-{\cal{A}}_3\,=\,\frac{1}{2}(q_{h_{u}}+q_{h_{d}}-4)\,=\,0\,\,mod(\eta)\\
\,\\
{\cal{A}}_1\,-{\cal{A}}_3\,=\,\frac{3}{10}(q_{h_{u}}+q_{h_{d}}-4)\,-\frac{12}{5}\,=\,0\,\,mod(\eta)
\end{array}{\label{NOANOM}}
\end{equation}
Imposing as a constraint the existence of the Yukawa couplings necessary for the fermion masses\footnote{These can be {\textit{``solved"}} $mod(N/2)$ as
$$q_{10}=1-q_{h_{u}}/2,\,\,\,q_{\overline{5}}\,=\,1\,+\,q_{h_{u}}/2\,-q_{h_{d}},\,\,\,q_1\,=\,1\,-3q_{h_{u}}/2\,+\,q_{h_{d}}\,.$$
Any allowed sets of charges should obey these relations.},
\begin{equation}2q_{10}\,+\,q_{h_{u}}\,=\,q_{10}\,+\,q_{\overline{5}}\,+\,q_{h_{d}}\,=\,q_{\overline{5}}\,+\,q_1\,+\,q_{h_{u}}\,=\,2\,\,\,mod(N),\,{\label{MASS}}\end{equation}
the conditions ({\ref{NOANOM}}) are equivalent to
\begin{equation}\begin{array}{l}
3\,-2N_f\,=\,0\,\,mod(\eta)\\
\,\\
(q_{h_{u}}+q_{h_{d}}-4)\,=\,0\,\,mod(N)\\
\,\\
12\,=\,0\,\,mod(\eta)
\end{array}\end{equation}
For $N_f=3$, this restricts the possible ${\cal{Z}}_N$ $R$-symmetries to
\begin{equation}N\,=\,3,\,6\,.\end{equation}

Note that the $\mu$-term is absent\footnote{Here, as in all subsequent cases, it is assumed that a $\mu$-term of the correct order of magnitude is generated through non-perturbative effects\cite{GM} or through another indirect mechanism\cite{KN}.}, since $q_u+q_{d}\,=\,4\,\neq\,2\,\,\,mod(N)\,.$ Dangerous $D=\,5$ Baryon/Lepton violating operators are also absent, since $QQQ\ell,\,u^cu^ce^cd^c\,\rightarrow\,3q_{10}\,+\,q_{\overline{5}}\,=\,4\,-q_{h_{u}}-q_{h_{d}}\,=\,0\,\,\,mod(N)\,$ and $u^cd^cd^cN^c\,\rightarrow\,2q_{\overline{5}}\,+\,q_{10}\,+\,q_1\,=\,4-q_{h_{u}}-q_{h_{d}}\,=\,0\,\,mod(N)$. Nevertheless, the potentially dangerous $D=3,\,4$ operators transform as $\ell\,h_u\rightarrow\,q_{\overline{5}}+q_{h_{u}},\,\,\,\,\,Q\ell\,d^c,\,d^cd^cu^c\rightarrow\,q_{10}+2q_{\overline{5}}\,$
 and a choice of charge assignments is required.

 In the $N=3$ case, since $q_{h_{u}}+q_{h_{d}}\,=\,1\,\,mod(3)$, for $q_{h_{u}}=0$ and $q_{h_{d}}=1$, we can have $q_{10}=1,\,q_{\overline{5}}=0$ and $q_1=2$. Then, the $D=3, 4$ matter-parity violating operators cannot be present. Note however that the right-handed neutrino Majorana mass is not possible. The vanishing of the coefficient ${\cal{A}}_1$ can be met for the shifted charges $\tilde{q}_{h_u}=12,\,\tilde{q}_{h_d}=10,\,\tilde{q}_{10}=4$.

 In the $N=6$ case, we have $q_{h_{u}}\,+\,q_{h_{d}}\,=\,4\,\,\,mod(6)$, which can be met only with even charges due to the mass relations ({\ref{MASS}}). The choice $q_{h_d}\,=\,q_{h_u}\,=\,2$ is $SO(10)$ invariant. The phenomenologically acceptable charges, for which the above $D=3,\,4$ operators are absent, are listed in the following table (Note that in Table-1 $\tilde{q}_a$ stands for shifted charges and the last entry refers to the right-handed neutrino Majorana mass operator.)$$\,$$
 {\centerline{\begin{tabular}{|c|c|c|c|c|c|c|c|}
\hline
 $q_{h_u}$ & $q_{h_d}$& $q_{10}$ & $q_{\overline{5}}$&$q_1$ & $\tilde{q}_{h_u}$& $\tilde{q}_{h_d}$ & $({N^c})^2$ \\
\hline
0 & 4 & 1 & 3 & 5 & 12 & 10 & $-$ \\
\hline
0 & 4 & 4 & 0 & 2 & 12 & 10 & $-$\\
\hline
4 & 0 & 5 & 3 & 1 & 10 & 12 & $+$\\
\hline
 2 & 2 & 3 & 3 & 3 & 14 & 8 & $-$\\
\hline
\end{tabular}}
\newline\newline
{\centerline{\textbf{Table 1: Allowed $\bf{SU(5)}$-Invariant Charges for ${\bf{{\cal{Z}}_6^{(R)}}}$}}}
\newline

In the above search for a non-anomalous ${\cal{Z}}_N$ $R$-symmetry it has been assumed that it commutes with $SU(5)$\cite{EIKY}. This need not be necessarily the case. Let's consider the following {\textit{``flipped"}} assignments
\begin{equation}q_Q\,=\,q_{d^c}\,=\,q_{N^c}\,=\,q_{10},\,\,\,q_{\ell}\,=\,q_{u^c}\,=\,q_{\overline{5}},\,\,\,q_{e^c}\,=\,q_1\,,\end{equation}
which commute with the so-called {\textit{``flipped"}} $SU(5)\times U(1)$. The corresponding gauge anomaly coefficients are identical to those of ({\ref{ANCOEF}}) with the exception of
\begin{equation}{\cal{A}}_1\,=\,\frac{N_f}{10}\left(\,3q_{10}+11q_{\overline{5}}+6q_1\,-20\right)\,+\,\frac{3}{10}(q_{h_{u}}+q_{h_{d}}-2)\,.\end{equation}
Note however that, when we impose the mass relations
\begin{equation}
2q_{10}+q_{h_{d}}\,=\,q_{10}+q_{\overline{5}}+q_{h_{u}}\,=\,q_{\overline{5}}+q_1+q_{h_{d}}\,=\,2\,\,\,mod(N)\,,{\label{FMASS}}
\end{equation}
 we can prove\footnote{These are now {\textit{``solved"}} by $q_{10}\,=\,1-q_{h_{d}}/2$,\,$q_{\overline{5}}\,=\,1+q_{h_{d}}/2-q_{h_{u}}$ and $q_1\,=\,1-3q_{h_{d}}/2+q_{h_{u}}$.}
$$3q_{10}\,+\,11q_{\overline{5}}\,+\,6q_1\,=\,5\left(3q_{10}\,+\,q_{\overline{5}}\right)\,,\,$$
leading to
\begin{equation}{\cal{A}}_1\,=\,\frac{N_f}{2}\left(\,3q_{10}+q_{\overline{5}}-4\right)\,+\,\frac{3}{10}(q_{h_{u}}+q_{h_{d}}-2)\,,\,\end{equation}
which is identical to the corresponding expression in ({\ref{ANCOEF}}). Thus, in both cases, of straight and flipped charge assignments, all gauge anomaly coefficients are identical, provided we invoke the conditions for the existence of mass Yukawa couplings. Nevertheless, it should be reminded that the anomaly cancellation condition on ${\cal{A}}_1$ rests on the existence of at least one set of shifted charges that satisfy the corresponding condition. Note also, even
in the case of non-unified assignments that will follow, we assume that the $U(1)_Y$ hypercharge factor is normalized in the standard $E_6$ fashion.

The allowed cases for a non-anomalous discrete $R$-symmetry are still $N\,=\,3,\,6$. Nevertheless, the resulting models need not be the same. The $\mu$-term is still absent, since $q_{h_{u}}\,+\,q_{h_{d}}\,=\,4\,\neq\,2\,\,\,mod(N)$. Similarly, for the $D=5$ Baryon/Lepton violating operators, we have $QQQ\ell,\,d^cd^cN^cu^c\rightarrow\,3q_{10}+q_{\overline{5}}\,=\,4-(q_{h_{u}}+q_{h_{d}})\,=\,0\,\neq\,2\,\,\,mod(N)$ and $u^cu^cd^ce^c\rightarrow\,2q_{\overline{5}}+q_{10}+q_1\,=\,4\,-(q_{h_{u}}+q_{h_{d}})\,=\,0\,\neq\,2\,\,\,mod(N)$. Thus, these operators are absent. In order to conclude whether the $D=3,\,4$ dangerous operators are allowed, we must proceed further with the charge assignments for each value of $N$.

In the $N=3$ case, we can take $q_{h_{d}}=1,\,q_{h_{u}}=0$. Then, we can have $q_{10}=2,\,q_{\overline{5}}\,=\,0,\,\,\,q_1\,=\,1$. The $D=3$ operator $\ell h_u\rightarrow\,q_{\overline{5}}+q_{h_{u}}\,=\,0$ cannot be present. The $D=4$ operators $Q\ell d^c,\,d^cd^cu^c\rightarrow\,2q_{10}+q_{\overline{5}}\,=\,1$ cannot be present either. Thus, the ${\cal{Z}}_3$ $R$-symmetry in this case of flipped assignments is also phenomenologically feasible. The $N=6$ case proceeds in an analogous fashion.
Taking $q_{h_d}=4,\,q_{h_u}=0$, we are led to $q_{10}=5,\,q_{\overline{5}}=3,\,q_1=1$, which disallows $\ell h_{u}$ and $Q\ell d^c,\,d^cd^cu^c$. Similarly, for the rest of the flipped charge assignments shown in Table 2. Shifted charges are denoted as $\tilde{q}$.

$$\,$$
{\centerline{\begin{tabular}{|c|c|c|c|c|c|c|c|c|c|c|c|}
\hline
${\cal{Z}}_N$&$q_{Q}$&$q_{d^c}$&$q_{N^c}$&$q_{\ell}$&$q_{u^c}$&$q_{e^c}$&$q_{h_{u}}$&$q_{h_{d}}$&$\tilde{q}_{e^c}$&$\tilde{q}_{h_{u}}$&$\tilde{q}_{h_d}$\\
\hline
$N=3$&$2$&$2$&$2$&$0$&$0$&$1$&$0$&$1$&$4$&$12$&$10$\\
\hline
$N=6$&$5$&$5$&$5$&$3$&$3$&$1$&$0$&$4$&$7$&$12$&$10$\\
\hline
$N=6$&$2$&$2$&$2$&$0$&$0$&$4$&$0$&$4$&$4$&$12$&$10$\\
\hline
$N=6$&$1$&$1$&$1$&$3$&$3$&$5$&$4$&$0$&$-1$&$10$&$12$\\
\hline
$N=6$&$3$&$3$&$3$&$3$&$3$&$3$&$2$&$2$&$3$&$14$&$8$\\
\hline
\end{tabular}}}
\newline\newline
{\centerline{\textbf{Table 2: Allowed Flipped Charges for ${\bf{{\cal{Z}}_3^{(R)}}}$ and ${\bf{{\cal{Z}}_6^{(R)}}}$}}}
\,

If the model is extended beyond $MSSM$ by the introduction of a singlet $S$ coupled to the Higgses through a term $Sh_uh_d$, the required charge of the new field has to be $q_S\,=\,2\,-q_{h_{u}}-q_{h_{d}}\,=\,-2\,\,\,mod(N)$, i.e. $q_S=1$ for ${\cal{Z}}_3$ or $q_S=4$ for ${\cal{Z}}_6$. Nevertheless, in both cases the term $S^2$ is allowed, corresponding to a large mass for the singlet and making this extension phenomenologically uninteresting.

{\textbf{$\bf{\tilde{N}\,>\,1}$ pairs of isodoublets.}} Another possible extension of the MSSM to be considered is the case of extra Higgs isodoublets with identical charges. The anomaly coefficients for $\tilde{N}\geq 1$ pairs of isodoublets take the form

\begin{equation}\begin{array}{l}
{\cal{A}}_3\,=\,3\,+\,\frac{N_f}{2}\left(\,3q_{10}\,+\,q_{\overline{5}}\,-4\right)\\
\,\\
{\cal{A}}_2\,=\,2\,+\,\frac{N_f}{2}\left(\,3q_{10}\,+\,q_{\overline{5}}\,-4\right)\,+\,\frac{\tilde{N}}{2}(q_{h_{u}}+q_{h_{d}}\,-2)\\
\,\\
{\cal{A}}_1\,=\,\frac{N_f}{2}\left(\,3q_{10}\,+\,q_{\overline{5}}\,-4\right)\,+\,\frac{3\tilde{N}}{10}(q_{h_{u}}+q_{h_{d}}-2)
\end{array}
\end{equation}

The corresponding conditions ${\cal{A}}_i=0\,\,\,mod(\eta)$ in the case $N_f\,=\,3$, after we enforce the, common in flipped or $SU(5)$ assignments, mass relation $3q_{10}+q_{\overline{5}}-4\,=\,-(q_{h_u}+q_{h_d})\,\,\,mod(N)$, amount to
\begin{equation}
\begin{array}{l}
{\cal{A}}_3\,\rightarrow \,3(q_{h_u}+q_{h_d}-2)\,=\,0\,\,mod(N)\\
\,\\
{\cal{A}}_2\,-{\cal{A}}_3\,\rightarrow\,\tilde{N}(q_{h_u}+q_{h_d}-2)\,=\,2\,\,\,mod(N)\\
\,\\
{\cal{A}}_1\,-{\cal{A}}_3\,\rightarrow\,\frac{3\tilde{N}}{5}(q_{h_u}+q_{h_d}-2)\,=\,6\,\,\,mod(N)\,\leftrightarrow\, 24\,=\,0\,\,\,mod(N)
\end{array}
\end{equation}
From the first two equations we obtain in the familiar case $\tilde{N}=1$ the standard non-trivial solutions $N=\,3,\,6$ with  $q_{h_u}+q_{h_d}\,=\,4\,\,mod(N)$ discussed previously. In an analogous manner, for $\tilde{N}=2$ we again obtain  $N=\,3,\,6$ for the allowed symmetries but with a different relation for the  Higgs charges, namely $q_{h_u}+q_{h_d}=\,0\,\,\,mod(N)$. In contrast, the $\tilde{N}=3$ case has no solution for non-trivial symmetries ($N\geq 3$).

{\section{Discrete $R$-Symmetries with Anomaly Cancellation through the GS Mechanism} Another possibility for the realization of discrete $R$-symmetries is when gauge and gravitational anomalies are cancelled through the operation of the {\textit{Green-Schwarz mechanism}}. The corresponding conditions on the anomaly coefficients ({\ref{ANCOEF}}) read
\begin{equation}
\begin{array}{l}
\mathcal{A}_3\,=\,\mathcal{A}_2\,=\,\mathcal{A}_1\,=\,\rho\,\,mod(\eta)\\
\,\\\
\mathcal{A}_0\,=\,24\rho\,\,mod(\eta)
\end{array}
{\label{COND}}\end{equation}
From the explicit expressions of the coefficients ({\ref{ANCOEF}}) for the $MSSM$ content, enforcing the mass conditions, we obtain\footnote{We have included in ${\cal{A}}_0$ the contribution of the right-handed neutrino and that of the dilatino/axino. Enforcing the mass conditions on it we obtain ${\cal{A}}_0\,=\,-14\,-10(q_{h_{u}}+q_{h_{d}})\,+\,\sum_s(q_s-1)$. The difference with ${\cal{A}}_3=3-\frac{3}{2}(q_{h_{u}}+q_{h_{d}})$ is ${\cal{A}}_0\,-24{\cal{A}}_3\,=\,18\,+\,26(q_{h_{u}}+q_{h_{d}}-4)\,+\,\sum_s(q_s-1)$.} for ${\cal{Z}}_N$ charges commuting with $SU(5)$ or flipped.
\begin{equation}
\begin{array}{l}
{\mathcal{A}}_3\,=\,\rho\,\rightarrow \rho\,=\,-3\,\,mod(\eta)\\
\,\\
\mathcal{A}_2-\mathcal{A}_3\,=\,0\,\rightarrow q_{h_u}\,+\,q_{h_d}\,=\,4\,\,mod(N)\\
\,\\
\mathcal{A}_1-\mathcal{A}_3\,=\,0\,\rightarrow\,\, 12\,=\,0\,\,mod(\eta)\\
\,\\
\mathcal{A}_0-24\mathcal{A}_3\,=\,0\,\rightarrow \,18\,+\,\sum_s(q_s-1)\,=\,0\,\,mod(\eta)\\
\,\\
\end{array}
{\label{COND-1}}\end{equation}
Note that the first three of these conditions, for $\rho\,=\,0$ (i.e. $N=3,\,6$), coincide with the conditions ({\ref{NOANOM}}).

The last two conditions can be reconciled without the use of extra singlets in the case of $N=3,\,4,\,6,\,12$. Nevertheless, they can be compatible for all possible $N$ if we extend $MSSM$ introducing extra singlets. A most straightforward possibility is that of a singlet $S$ coupled to the Higgses as $S\,h_u\,h_{d}$. Then, its required charge would be $q_S\,=\,-2\,\,\,mod(N)$. Two such singlets reconcile the last two conditions\footnote{Of course, such a ``solution" is not unique. Six singlets of zero charge could do the job as well. Note also that, in the case of $N=3,\,6$, these conditions are met with an arbitrary number of such singlets, since $\sum_s(q_S-1)\,=\,N_s(-3\,\,mod(N)\,)$, which is a subset of $0\,\,mod(3)$.}. Thus, the allowed discrete $R$-symmetries would be those corresponding to $N\,=3,\,4,\,6,\,8,\,12,\,24$. Note that in the case $N=3,\,6$ only the quadratic term $S^2$ is allowed. In the $N=4$ case, only a linear and a cubic term are allowed. Finally, in the $N=8$ case, only a cubic term is allowed. Thus, in the $N=8$ case the singlet extension coincided with $NMSSM$, in the $N=4$ case it is a modified $NMSSM$ with an additional linear term. Finally, the cases $N=12,\,24$, where no singlet self-term is present, correspond to what has been termed $nMSSM$\cite{PT}.

Allowed ${\cal{Z}}_N$ $R$-symmetries commuting with $SU(5)$ or $SO(10)$ have been studied\cite{LRRRSHV}. We shall go further in our analysis considering charge assignments that do not commute with these symmetries. More specifically, we shall consider {\textit{``flipped"}} assignments
\begin{equation}q_Q\,=\,q_{d^c}\,=\,q_{N^c}\,=\,q_{10},\,\,\,q_{\ell}\,=\,q_{u^c}\,=\,q_{\overline{5}},\,\,\,q_{e^c}\,=\,q_1\,,\end{equation}
which commute with the so-called {\textit{``flipped"}} $SU(5)\times U(1)$. The anomaly coefficient conditions considered above are identical for all these cases, provided the mass relations are invoked. Nevertheless, the final charge assignments for matter correspond to distinct models. Turning now to the phenomenological features of the allowed models, we see that the condition on the Higgs charges ({\ref{COND-1}}) is sufficient to forbid the $\mu$-term. In addition, dangerous $D=5$ Baryon/Lepton Number violating operators are absent as well, due to ({\ref{FMASS}}) and ({\ref{COND-1}}). Indeed, we have
$$QQQ\ell,\,\,d^cd^cu^cN^c\,\,\rightarrow\,\,3q_{10}\,+\,q_{\overline{5}}\,=\,4\,-q_{h_u}-q_{h_{d}}\,=\,0\,\neq\,2\,\,\,mod(N)\,$$
$$u^cu^cd^ce^c\,\,\rightarrow\,2q_{\overline{5}}\,+\,q_{10}\,+\,q_1\,=\,4\,-q_{h_u}-q_{h_{d}}\,=\,0\,\neq\,2\,\,\,mod(N)\,.$$
In order to conclude whether dangerous $D=3,\,4$ operators are present we need to proceed further with the matter and Higgs charge assignments for each particular value of $N$.

$\bf{N\,=\,3,\,6}$. In these cases the first of the conditions ({\ref{COND-1}}) implies that $\rho\,=\,0$. Thus, for these cases the analysis will coincide with the one carried on previously in anomaly-free case without the operation of the GS mechanism.

$\bf{N=4}$. The condition $q_{h_{u}}+q_{h_{d}}\,=\,0\,\,\,mod(4)$ can only be satisfied with even charges. These are the choices $q_{h_{u}}=q_{h_{d}}=0$ and $q_{h_{u}}=q_{h_{d}}=2$. The first choice leads to two $SO(10)$-invariant solutions with identical phenomenology. Note that these solutions allow for a right-handed neutrino Majorana mass. All constraints for the absence of $D\,=\,3,\,4$ operators are met. The second choice  is discarded due to the presence of unwanted $D=3,4$ operators. The allowed cases are listed in Table-3.
$$\,$$
{\centerline{\begin{tabular}{|c|c|c|c|c|c|c|c|c|c|c|}
\hline
$N$&$q_{Q}$&$q_{d^c}$&$q_{N^c}$&$q_{\ell}$&$q_{u^c}$&$q_{e^c}$&$q_{h_{u}}$&$q_{h_{d}}$&$\tilde{q}_{h_{u}}$&$\tilde{q}_{h_{d}}$\\
\hline
4&1&1&1&1&1&1&0&0&4&8\\
\hline
4&3&3&3&3&3&3&0&0&4&8\\
\hline
\end{tabular}}}
\newline
{\centerline{\textbf{Table 3: Allowed Flipped Charges for ${\bf{{\cal{Z}}_4^{(R)}}}$}}}
\newline

$\bf{N\,=\,8,\,12,\,24}$. The condition $q_{h_{u}}+q_{h_{d}}\,=\,4\,\,\,mod(N)$ allows again only even charges and can in principle be satisfied for various sets of Higgs charges. Among them two distinct general solutions, with $D=3,4$ operators absent, are listed in Table-4, where, of course, these charges can always be modulo-shifted to lie in the first modulo of the given symmetry. The charge $\tilde{q}_{N^c}=1$ of the second row allows always for a right-handed neutrino Majorana mass term.
\newline\newline
{\centerline{\begin{tabular}{|c|c|c|c|c|c|c|c|c|}
\hline
$N$&$\tilde{q}_{Q}$&$\tilde{q}_{d^c}$&$\tilde{q}_{N^c}$&$\tilde{q}_{\ell}$&$\tilde{q}_{u^c}$&$\tilde{q}_{e^c}$&$\tilde{q}_{h_{u}}$&$\tilde{q}_{h_{d}}$\\
\hline
8,12,24&-1&-1&-1&3&3&-5&24&28\\
\hline
8,12,24&1&1&1&-3&-3&5&28&24\\
\hline
\end{tabular}}}
\newline
{\centerline{\textbf{Table 4: Allowed (Shifted) Flipped Charges for ${\bf{{\cal{Z}}_8^{(R)}}},\,{\bf{{\cal{Z}}_{12}^{(R)}}},\,{\bf{{\cal{Z}}_{24}^{(R)}}}$}}}
$$\,$$

{\textbf{A ``non-unified" case ($\bf{{\cal{Z}}_5^{(R)}}$).}} The above found set of allowed ${\cal{Z}}_N$ does not exclude other {\textit{``non-unified"}} possibilities. The anomaly coefficients for non-unified charges are
\begin{equation}
\begin{array}{l}
{\cal{A}}_3\,=\,3\left(q_Q+\frac{1}{2}q_{d^c}+\frac{1}{2}q_{u^c}-2\right)\,+\,3\\
\,\\
{\cal{A}}_2\,=\,3\left(\frac{3}{2}q_Q+\frac{1}{2}q_{\ell}-2\right)\,+\,\frac{\tilde{N}}{2}\left(\,q_{h_d}+q_{h_u}\,-2\right)\,+\,2\\
\,\\
{\cal{A}}_1\,=\,\frac{9}{5}\left(\frac{1}{6}q_Q+\frac{1}{3}q_{d^c}+\frac{4}{3}q_{u^c}+\frac{1}{2}q_{\ell}+q_{e^c}-\frac{10}{3}\right)\,+\,\frac{3\tilde{N}}{10}(q_{h_d}+q_{h_u}-2)
\end{array}
\end{equation}
where $\tilde{N}$ is the number of Higgs isodoublets. A particularly interesting case is that of a ${\cal{Z}}_5$ discrete $R$-symmetry, motivated by the solution to the $\mu$-problem in terms of a minimal singlet extension of MSSM. The assigned charges\cite{PT} are
$$\,$$
{\centerline{\begin{tabular}{|c|c|c|c|c|c|c|c|c|c|}
\hline
$N=5$&$q_{Q}$&$q_{\ell}$&$q_{d^c}$&$q_{u^c}$&$q_{e^c}$&$q_{N^c}$&$q_{S}$&$q_{h_{u}}$&$q_{h_{d}}$\\
\hline
&4&4&1&1&1&1&3&2&2\\
\hline
\end{tabular}}}

{\centerline{\textbf{Table 5: $\bf{{\cal{Z}}_5^{(R)}}$ Charges}}}

For these charges the anomaly coefficients are
\begin{equation}
\begin{array}{l}
{\cal{A}}_3\,=\,\rho\,=\,2\,\,mod(5)\\
\,\\
{\cal{A}}_2\,=\,\tilde{N}\,\,mod(5)\\
\,\\
{\cal{A}}_1\,=\,\frac{24}{5}\,+\,\frac{3N_{\sigma}}{5}(q_{\sigma}+q_{\overline{\sigma}}-2)\,\,mod(5)\\
\end{array}{\label{KONDO-1}}
\end{equation}
where we have introduced tentatively $N_{\sigma}$ pairs of hypercharge $\pm 1$ singlets. Finally, the condition on ${\cal{A}}_0\,=\,70\,+\,N_{\sigma}(q_{\sigma}+q_{\overline{\sigma}}-2)\,+\,\sum_s(q_s-1)$ reads $2\,+\,N_{\sigma}(q_{\sigma}+q_{\overline{\sigma}}-2)\,+\,\sum_s(q_s-1)\,=\,0\,\,mod(5)$. The anomaly conditions corresponding to ({\ref{KONDO-1}}) can be met for $\tilde{N}=2$, $N_{\sigma}=1$ and the shifted charges\footnote{Another working choice is $N_{\sigma}=2$ and $q_{\sigma}\,=\,q_{\overline{\sigma}}=4$.} $q_{\sigma}=q_{\overline{\sigma}}=7$. The gravitational anomaly condition can also be met if, apart from the charge-3 singlet $S$, we introduce also an additional neutral singlet. Note that for these charges all possible dangerous terms are disallowed ($\sigma\overline{\sigma},\,e^c\sigma,\,\ell\overline{\sigma}h_d,\,h_{d}d_{d}\overline{\sigma},\,h_uh_u\sigma$). Note also that a term $\sigma\,\overline{\sigma}\,S$ is allowed. Apart from being anomaly-free, this model allows for all standard terms, including neutrino Majorana masses, and forbids all unwanted $D\leq 5$ terms. The extra charged singlets introduced here for the sake of anomaly cancellation can obtain a mass through the vev of the singlet $S$.

\section{Unification} Can these discrete symmetries be incorporated in a $4D$ grand unified theory? There are convincing arguments that, for a simple gauge group and $MSSM$ particle content at low energies, this is not possible\cite{FRV}. We shall depart from both of these assumptions and allow on the one hand, additional matter at low energies, beyond $MSSM$, and on the other hand go beyond simple unifying groups. Specifically, we shall consider $SU(5)\times U(1)$ and use {\textit{``flipped"}} ${\cal{Z}}_N$-charge assignments commuting with it. One of the motivations for this model is that it is accompanied by an elegant mechanism for {\textit{triplet-doublet splitting}}. Note that this is one of the main problems that one has to face in promoting $MSSM$ to a GUT endowed with the discussed discrete $R$-symmetries.

{\textbf{Minimal Case.}} The standard matter content of $SU(5)\times U(1)$ comes in three copies of $F(10,\,1),\,f^c({\overline{5}},\,-3),\,\ell^c(1,\,5)$ with corresponding ${\cal{Z}}_N$ charges $q_{10},\,q_{\overline{5}},\,q_1$. The standard Higgs content is $h(5,\,-2;\,{q_{h}}),\,h^c(\overline{5},\,2;\,{q_{h^{c}}}),\,H(10,\,1;\,0),\,\overline{H}(\overline{10},\,-1;\,0)$. The last pair, through its non-zero vev will achieve the breaking down to $MSSM$. Obviously, it has to be neutral under the discrete symmetry. The standard matter couplings are
\begin{equation}F\,F\,h\,+\,F\,f^c\,h^c\,+\,f^c\,\ell^c\,h^c\,.\end{equation}
Nevertheless, in order to realize triplet-doublet splitting, the couplings $H\,H\,h\,+\,\overline{H}\,\overline{H}\,h^c$ are necessary. These couplings force the Higgs charges
\begin{equation}q_h\,=\,q_{h^c}\,=\,2\,\,\,mod(N)\,.{\label{EQ}}\end{equation}
Note that these charges do not allow a $\mu$-term. Nevertheless, enforcing the mass conditions
\begin{equation}
2\,q_{10}\,+\,q_h\,=\,q_{10}\,+\,q_{\overline{5}}\,+\,q_{h^c}\,=\,q_{\overline{5}}\,+\,q_1\,+\,q_{h}\,=\,2\,\,mod(N)\,,\,{\label{KONDI-0}}\end{equation}
 we are led to the $SO(10)$-invariant assignments
\begin{equation}q_{10}\,=\,q_{\overline{5}}\,=\,q_1\,=\,\frac{N}{2}\,\,\,mod(N)\,\,\,\,\,\,\,\,\,(N\,>\,4)\,.\end{equation}
Note that, since the charges have to be integers, only even values of
$N$ are allowed. Note also that the case $N=4$, in which all matter fields have charge $2\,\,mod(4)$, is excluded since it allows the operator $F\overline{H}$.
The anomaly coefficient conditions for general even $N\,>\,4$ are
\begin{equation}\begin{array}{l}
{\cal{A}}_5\,=\,3N-3\,=\,\rho\,\,\,mod(N/2)\,\\
\,\\
{\cal{A}}_X\,=\,3N\,-3\,-\frac{5}{2}\,=\,\rho\,\,\,mod(N/2)
\end{array}\end{equation}
and anomaly cancellation cannot be met without introducing extra matter. It can also be checked that anomaly cancellation cannot be saved by shifted charges.

{\textbf{Extra Fields.}} In what follows we shall introduce extra matter and at the same time avoid the $SO(10)$-symmetric assignment enforced by ({\ref{EQ}}) in order to realize the split assignments found in the previous section. In order to avoid ({\ref{EQ}}), we introduce an extra pair of Higgs fields $H'(10,\,1;\,q_{H'}),\,\,\,\overline{H}^{\,\prime}(\overline{10},\,-1;\,q_{\overline{H}^{\,\prime}})$.
Assuming that the only non-zero vevs are those\footnote{Thus, assuming a ${\cal{Z}}_N^{(R)}$-invariant vacuum.} of $\langle H\rangle=\langle\overline{H}\rangle$ and assuming that the couplings $H\,H'\,h,\,\,\overline{H}\,\overline{H}^{\,\prime}\,h^c$
are present, after symmetry breaking, the pairs $d_h(3,\,1,\,-1/3;\,q_h),\,\,\,\,d_{H'}^{c}(\overline{3},\,1,\,1/3;\,q_{H'})$ and $d_{h^c}^c(\overline{3},\,1,\,1/3;\,q_{h^c}),\,\,\,\,d_{\overline{H}^{\,\prime}}({3},\,1,\,-1/3;\,q_{\overline{H}^{\,\prime}})$
will obtain a large mass and will be removed from the spectrum. In addition, the fields $Q_H,\,Q^c_{\overline{H}},\,N_H^c-N^c_{\overline{H}}$ will be {\textit{``higgsed away"}}. The surviving fields, apart from the standard matter and a neutral Higgs singlet, will be a {\textit{``hybrid"}} pair $10'+\overline{10}\,'$ composed out of the $H'(\overline{H}^{\,\prime})$ with their $d^c_{H'}(d_{\overline{H}^{\,\prime}})$ replaced by $d^c_H(d_{\overline{H}})$. The conditions on the charges resulting from the existence of the standard couplings
\begin{equation}F\,F\,h\,+\,F\,f^c\,h^c\,+\,f^c\,\ell^c\,h^c\,+\,H\,H'\,h\,+\,\overline{H}\,\overline{H}^{\,\prime}\,h^c\end{equation}
are given by ({\ref{KONDI-0}}) and
\begin{equation}
\,q_{H'}\,+\,q_h\,=\,q_{\overline{H}^{\,\prime}}\,+\,q_{h^c}\,=\,2\,\,mod(N)\,.{\label{KONDI}}\end{equation}
Nevertheless, a number of unwanted terms are still allowed by gauge symmetry, namely, the terms
$$h\,h^c\,+\,H'\,\overline{H}\,+\,\overline{H}^{\,\prime}\,H\,+\,\overline{H}^{\,\prime}\,H'\,+\,F\,\overline{H}\,+\,F\,\overline{H}^{\,\prime}$$
\begin{equation}\,+\,H^2\,h\,+\,\overline{H}^2\,h^c\,+\,F\,H\,h\,+\,F\,H'\,h\,+\,H\,f^c\,h^c\,+\,H'\,f^c\,h^c\,,\,{\label{DANG}}\end{equation}
which are expected to be removed by the discrete symmetry.

Next let's consider the anomaly coefficients. The gauge anomaly coefficients are
\begin{equation}
\begin{array}{l}
{\cal{A}}_5\,=\,5\,+\,3\left(\frac{3}{2}q_{10}+\frac{1}{2}q_{\overline{5}}-2\right)\,+\,\frac{3}{2}\left(\,q_{H'}+q_{\overline{H}^{\,\prime}}\,-4\right)\,+\,\frac{1}{2}\left(q_h+q_{h^c}\,-2\right)\\
\,\\
{\cal{A}}_X\,=\,\frac{3}{40}\left(\,10q_{10}+45q_{\overline{5}}+25q_1-80\right)\,+\,\frac{1}{4}\left(q_{H'}+q_{\overline{H}^{\,\prime}}-4\right)\,+\,\frac{1}{2}(q_h+q_{h^c}-2)
\end{array}
\end{equation}
or, applying the conditions ({\ref{KONDI-0}}), ({\ref{KONDI}}),
\begin{equation}\begin{array}{l}
{\cal{A}}_5\,=\,4\,-\frac{5}{2}(q_h+q_{h^c})\,=\,\rho\,\,\,mod(\eta)\\
\,\\
{\cal{A}}_X\,=\,-1\,-\frac{5}{4}(q_h+q_{h^c})\,=\,\rho\,\,mod(\eta)
\end{array}
\end{equation}
Taking the familiar condition
\begin{equation}q_h+q_{h^c}\,=\,4\,\,\,mod(N)\,,{\label{HIGGS}}\end{equation}
 we obtain
 \begin{equation}
 {\cal{A}}_5\,=\,{\cal{A}}_X\,=\,-6\end{equation}
 which satisfies the anomaly condition for any $N$.

The gravitational coefficient is\footnote{The contribution of a dilatino/axino is also considered.} ${\cal{A}}_0\,=\,-\,7\,-\,17(q_h+q_{h^c})\,+\,\sum_s(q_s-1)$ and the corresponding condition becomes
\begin{equation}{\cal{A}}_0\,-24{\cal{A}}_5\,=\,69\,+\,\sum_s(q_s-1)\,=\,0\,\,\,mod(\eta)\end{equation}
and can be satisfied, among other choices, for any $N$ with a single neutral singlet of shifted charge $\tilde{q}_s=-68\,$. For all symmetries which are divisors of $138$ this extra singlet is not required.

Note that the $D=3,\,4$ Baryon/Lepton violating operators of the standard field content $\ell\,h_u,\,\,Q\ell d^c,\,u^cd^cd^c$, although not directly present due to the gauge symmetry, may appear through higher dimensional operators\footnote{$f^ch^c$, $FFf^c$ are not gauge singlets but $Hf^ch^c$, $HFFf^c$ are.}. Thus, their absence will be eventually determined by their respective ${\cal{Z}}_N$ charges. On the other hand, the analogous $D=5$ operators  $FFFf^c,\,\,\,f^cf^c{{F}}\ell^c$ are always absent since they have charges $3q_{10}\,+\,q_{\overline{5}}$ and $2q_{\overline{5}}+q_{10}+q_1$, both equal to $4-q_{h^c}-q_{h}\,=\,0\,\,\,mod(N)$, due to ({\ref{HIGGS}}).

Taking ({\ref{HIGGS}}) as our starting point, we proceed to consider phenomenologically allowed charge assignments. For even $N$, a general assignment that satisfies ({\ref{HIGGS}}) and is compatible with ({\ref{KONDI-0}}), ({\ref{KONDI}}) of {\textit{shifted charges}} is
\begin{equation}\begin{array}{l}
q_h\,=\,-2,\,q_{h^c}\,=\,6,\,q_{H'}\,=\,4,\,q_{\overline{H}^{\,\prime}}\,=\,-4\\
\,\\
\,q_{10}\,=\,2+N/2,\,q_{\overline{5}}\,=\,-6+N/2,
\,q_1\,=\,10\,+\,N/2
\end{array}\end{equation}
The anomaly constraints are readily satisfied\footnote{Note that the anomaly coefficients for the low energy spectrum with the new extra matter satisfy also ${\cal{A}}_3\,=\,{\cal{A}}_2\,=\,{\cal{A}}_1\,=\,3N-6$.} ${\cal{A}}_5\,=\,{\cal{A}}_X\,=\,3N\,-6\,.$
Next, we demand that the dangerous terms ({\ref{DANG}}) are absent. Their charges are
\begin{equation}\begin{array}{l}
hh^c\rightarrow 4,\,H'\overline{H}\rightarrow 4,\,\overline{H}^{\,\prime}H\rightarrow -4,\,\overline{H}^{\,\prime}H'\rightarrow 0\\
\,\\
F\overline{H}\rightarrow 2+N/2,\,F\overline{H}^{\,\prime}\rightarrow -2+N/2,\,H^2h\rightarrow\,-2,\,\overline{H}^2h^c\rightarrow\,6\\
\,\\
FHh\rightarrow\,N/2,\,FH'h\rightarrow 4+N/2,\,Hf^ch^c\rightarrow\,N/2,\,H'f^ch^c\rightarrow\,4+N/2
\end{array}\end{equation}
and they should be $\neq\,2\,\,\,mod(N)$. The first two rows lead to the constraints $N\,\neq\,4,\,6,\,8$, while the third does not give any additional constraint. Therefore, the allowed even values of $N$ are
\begin{equation}N\,=\,2k\,\geq\,10\,.\end{equation}
Note that, since $F^2$ has charge $4+N$, no right-handed neutrino Majorana mass is allowed. Furthermore the dangerous $D=3,\,4$ operators are absent since
\begin{equation}\begin{array}{l}
\ell h_u\rightarrow N/2\,,\,\,\,\,\,Q\ell d^c,\,d^cd^cu^c\rightarrow N/2-2+N
\end{array}\end{equation}
which are  $\neq2\,\,\,mod(N)$ for the allowed symmetries.
The corresponding charges for the cases $N=10,\,12$ are shown in Table-6.
\newline\newline
{\centerline{\begin{tabular}{|c|c|c|c|c|c|c|c|}
\hline
$N$&$q_{10}$&$q_{\overline{5}}$&$q_1$&$q_{h^c}$&$q_{h}$&$q_{H'}$&$q_{\overline{H}^{\,\prime}}$\\
\hline
10&7&9&5&6&8&4&6\\
\hline
12&8&0&4&6&10&4&8\\
\hline
\end{tabular}}}
\newline\newline
{\centerline{\textbf{Table 6. Allowed $\bf{{\cal{Z}}_{10}^{(R)},\,{\cal{Z}}_{12}^{(R)}}$ Charges for $\bf{SU(5)\times U(1)_X}$.}}}

Next, starting again from ({\ref{HIGGS}}), we proceed to investigate possible odd values of $N$ and consider a general assignment of {\textit{shifted}} charges compatible with ({\ref{KONDI-0}}), ({\ref{KONDI}})
\begin{equation}\begin{array}{l}
q_h\,=\,1,\,q_{h^c}\,=\,3,\,q_{H'}\,=\,1,\,q_{\overline{H}^{\,\prime}}\,=\,-1\\
\,\\
q_{10}\,=\,(N+1)/2,\,q_{\overline{5}}\,=\,(N-3)/2,\,q_1\,=\,(N+5)/2
\end{array}\end{equation}
The anomaly constraints are readily satisfied\footnote{Again, the anomaly coefficients for the low energy spectrum with the new extra matter satisfy also ${\cal{A}}_3\,=\,{\cal{A}}_2\,=\,{\cal{A}}_1\,=\,3N-6$.} ${\cal{A}}_5\,=\,{\cal{A}}_X\,=\,3N\,-6\,.$
Again, we demand that the dangerous terms ({\ref{DANG}}) are absent. Their charges are
\begin{equation}\begin{array}{l}
hh^c\rightarrow 4,\,H'\overline{H}\rightarrow 1,\,\overline{H}^{\,\prime}H\rightarrow -1,\,\overline{H}^{\,\prime}H'\rightarrow 0\\
\,\\
{F}\overline{H}\rightarrow (N+1)/2,\,{F}\overline{H}^{\,\prime}\rightarrow (N-1)/2,\,H^2h\rightarrow\,1,\,\overline{H}^2h^c\rightarrow\,3\\
\,\\
{F}Hh\rightarrow\,(N+3)/2,\,{F}H'h\rightarrow (N+5)/2,\,Hf^ch^c\rightarrow\,(N+3)/2,\,H'f^ch^c\rightarrow\,(N+5)/2
\end{array}\end{equation}
and they should be $\neq\,2\,\,\,mod(N)$. The first two rows lead to the constraints $N\,\neq\,3,\,5$, while the third does not supply us with any additional restriction. Thus, the allowed odd values of $N$ are
\begin{equation} N\,=\,2k+1\,\geq\,7\,.\end{equation}
Again, since $F^2$ has charge $N+1$, no right-handed neutrino Majorana mass is allowed. The $D=3,4$ operators are also absent.
In the Table-7 we show the corresponding charges for the cases $N=7,\,9$.
\newline\newline
{\centerline{\begin{tabular}{|c|c|c|c|c|c|c|c|}
\hline
$N$&$q_{10}$&$q_{\overline{5}}$&$q_1$&$q_{h^c}$&$q_{h}$&$q_{H'}$&$q_{\overline{H}^{\,\prime}}$\\
\hline
7&4&2&6&3&1&1&6\\
\hline
9&5&3&7&3&1&1&8\\
\hline
\end{tabular}}}
\newline\newline
{\centerline{\textbf{Table 7. Allowed $\bf{{\cal{Z}}_{7}^{(R)},\,{\cal{Z}}_{9}^{(R)}}$ Charges for $\bf{SU(5)\times U(1)_X}$.}}}

It can also be checked that for any other Higgs charge assignments  in the range $(0,7)$ satisfying $q_h+q_{h^c}=4\,\,\,mod(N)$, the corresponding phenomenologically viable models also forbid the symmetries ${\cal{Z}}_3,\,{\cal{Z}}_4,\,{\cal{Z}}_5,\,{\cal{Z}}_6,\,{\cal{Z}}_8$. This fact is sufficient to forbid these symmetries for all possible models with $q_h+q_{h^c}=4\,\,\,mod(N)$ although departing from this relation may in principle allow some of them. 

Before closing this section it is interesting to note that the above list of symmetries does not exhaust all possible symmetries for the given gauge group. As an example, consider the model of ref. \cite{DPT} that is characterized by a radiative breaking of the $SU(5)\times U(1)$ symmetry. This model, having the same set of fields as the model considered above plus gauge singlets, possesses the discrete symmetry ${\cal{Z}}_3^{(R)}\times {\cal{Z}}_2$, which can readily promoted to be anomaly-free at the expense of introducing a massive pair of hypercharge $\pm 1$ singlets of ${\cal{Z}}_3^{(R)}$-charge equal to $2$.

\section{Brief Conclusions} In the present article we have reconsidered the issue of possible anomaly-free discrete $R$-symmetries ${\cal{Z}}_N^{(R)}$ that avoid the $\mu$-problem and the dangerous $D\leq 5$ operators within MSSM and extensions of it. Freedom from anomalies was considered either through strictly vanishing anomaly coefficients for the (gauge)${}^2{\cal{Z}}_N$ anomalies or through the operations of the {\textit{Green-Schwarz mechanism}} for the former as well as the (grav)${}^2{\cal{Z}}_N$ anomalies. We have extended known investigations by considering charge assignments, that do not commute with the standard $SU(5)$ or $SO(10)$ gauge groups but are, instead, compatible with a so-called flipped-$SU(5)$ symmetry. Staying within the framework of $MSSM$, we have found  that, in the anomaly-free case, the symmetries ${\cal{Z}}_3^{(R)},\,{\cal{Z}}_6^{(R)}$ with flipped assignments are possible. We have also investigated the possibility of multidoublet extensions of MSSM in this case. Phenomenologically acceptable charge assignments have been listed. In the same framework, in the case of GS-anomaly cancellation, we have arrived at phenomenologically allowed ${\cal{Z}}_N^{(R)}$ symmetries with flipped assignments for $N\,=\,3,\,4,\,6,\,8,\,12,\,24$. Phenomenologically acceptable charges for this cases have been listed. As an example discrete symmetries non-commuting with any of the above unifying symmetries, we have also considered a ${\cal{Z}}_5^{(R)}$ symmetry with non-unified charges in the framework of a singlet extension of MSSM. We have also considered the question of finding such symmetries for traditional $4D$ grand unified models. Having excluded simple gauge groups such as $SU(5)$ or $SO(10)$ or non-simple groups like $SU(5)\times U(1)_X$ with MSSM-low energy content, we arrived at an extended flipped $SU(5)\times U(1)_X$ model. For this model, ${\cal{Z}}_N^{(R)}$ symmetries were shown to be anomaly free (through GS) and phenomenologically viable for $N\,=\,2k+7\,$ and $N\,=\,2k+10\,$.

{\textbf{Acknowledgements}}

This research has been co-financed by the European Union (European Social Fund – ESF) and Greek national funds through the Operational Program ``Education and Lifelong Learning" of the National Strategic Reference Framework (NSRF) - Research Funding Program: ``Heracleitus II. Investing in knowledge society through the European Social Fund". Both authors acknowledge the hospitality of the CERN Theory Group.


\begin{thebibliography}{99}

\bibitem{FF} G. R. Farrar and P. Fayet, Phys. Lett. {\textbf{B76}} (1978), 575.

\bibitem{DRW} S. Dimopoulos, S. Raby and F. Wilczek, Phys. Lett. {\textbf{B112}} (1982), 133; N.Sakai and T.Yanagida, Nucl. Phys. {\textbf{B197}} (1982) 533;
S.Weinberg, Phys. Rev. {\textbf{D26}} (1982) 287.

\bibitem{IR} L. E. Ibanez and G. G. Ross, Nucl. Phys. {\textbf{B368}} (1992), 3; Phys. Lett. {\textbf{B260}} (1991), 1221.

\bibitem{BGW} K. S. Babu, I. Gogoladze and K. Wang, Phys. Lett. {\textbf{B570}} (2003), 32.

\bibitem{DLT} H. K. Dreiner, C. Luhn and M. Thormeier, Phys. Rev. Lett. {\textbf{D73}} (2006), 075007.

\bibitem{PT} C. Panagiotakopoulos and K. Tamvakis, Phys.Lett. B446 (1999), 224; Phys.Lett. B469 (1999) 145;

A. Dedes, C. Hugonie, S. Moretti, K. Tamvakis, Phys.Rev. D63 (2001) 055009C.

Panagiotakopoulos and A. Pilaftsis, Phys.Rev. D63 (2001) 055003; Phys.Lett. B505 (2001) 184.


\bibitem{KW} L. M. Krauss and F. Wilczek, Phys. Rev. Lett. {\textbf{62}} (1989), 1221.

\bibitem{BD} T. Banks and M. Dine, Phys. Rev. {\textbf{D45}} (1992), 1424.

\bibitem{EIKY} Jason L. Evans, Masahiro Ibe, John Kehayias and Tsutomu T. Yanagida, arXiv:1111.2481 [hep-ph].

\bibitem{BGW-1} K. S. Babu, I. Gogoladze and K. Wang, Nucl. Phys. {\textbf{B660}} (2003), 322.

\bibitem{LRRRSHV} 	Hyun Min Lee, Stuart Raby, Michael Ratz, Graham G. Ross, Roland Schieren, Kai Schmidt-Hoberg and Patrick K.S. Vaudrevange,
Phys.Lett. {\textbf{B694}} (2011), 491; Nucl.Phys. {\textbf{B850}} (2011), 1.

\bibitem{GM} G. Giudice and A. Masiero, Phys. Lett. {\textbf{B206}} (1988), 480.

\bibitem{KN} J. E. Kim and H. P. Nilles, Phys. Lett. {\textbf{B138}} (1984), 150.


\bibitem{FRV} M. Fallbacher, M. Ratz and P. K. S. Vaudrevange, Phys.Lett. {\textbf{B705}} (2011), 503.

\bibitem{DPT} A. Dedes, C. Panagiotakopoulos and K. Tamvakis, Phys.Rev. {\textbf{D57}} (1998) 5493.


\end{thebibliography}
\end{document}